\documentclass[%
aip,
jmp,
 amsmath,amssymb,
preprint,%
]{revtex4-1}

\usepackage{graphicx}% Include figure files
\usepackage{dcolumn}% Align table columns on decimal point
\usepackage{bm}% bold math
\usepackage[utf8]{inputenc}
\usepackage[T1]{fontenc}
\usepackage{mathptmx}

\DeclareMathOperator{\arccosh}{arccosh}

\newcommand{\be}{\begin{equation}}
\newcommand{\ee}{\end{equation}}

\begin{document}

\title{The Wigner function of a semiconfined harmonic oscillator model with a position-dependent effective mass}

\author{S.M. Nagiyev}%
 \email{sh.nagiyev@physics.science.az}
\author{A.M. Jafarova}%
 \email{a.jafarova@physics.science.az}
\author{E.I. Jafarov}
\thanks{Corresponding author}
 \email{ejafarov@physics.science.az}
\affiliation{%
Institute of Physics, Ministry of Science and Education \\ Javid ave. 131, AZ1143, Baku, Azerbaijan%\\This line break forced% with \\
}%

\date{\today}% It is always \today, today,

\begin{abstract}

We propose a phase-space representation concept in terms of the Wigner function for a quantum harmonic oscillator model that exhibits the semiconfinement effect through its mass varying with the position. The new method is used to compute the Wigner distribution function exactly for such a semiconfinement quantum system. This method suppresses the divergence of the integrand in the definition of the quantum distribution function and leads to the computation of its analytical expressions for the stationary states of the semiconfined oscillator model. For this quantum system, both the presence and absence of the applied external homogenous field are studied. Obtained exact expressions of the Wigner distribution function are expressed through the Bessel function of the first kind and Laguerre polynomials. Furthermore, some of the special cases and limits are discussed in detail.

\end{abstract}

\maketitle

\section{Introduction}

The concept of phase space can be considered the best tool for the description of the dynamics of the mechanical system. The phase space of any classical mechanical system contains all possible values of its position and momentum as well as its time evolution through certain phase-space trajectories. The quantum world of similar dynamical systems is vastly different and extremely complex. Within the quantum approach, one deal with the probabilistic description of the sub-micron-sized physical systems through their non-commuting position and momentum operators. Then, the joint distribution of the momentum and position for quantum mechanical systems needs a new mathematical tool that can be as illuminating for us as in the case of classical mechanical systems. This potent mathematical tool is the Wigner distribution function~\cite{wigner1932}. It enables us to describe the quantum systems under study by using the language of classical physics~\cite{hillery1984}. 

There are numerous studies that deal with the computation of the Wigner function of various constant mass quantum harmonic oscillator models~\cite{davies1975,nagiyev1998,jafarov2007,jafarov2008,nagiyev2009,li2010,jafarov2010,kai2011,vanderjeugt2013,vanderjeugt2014,hassanabadi2016,hanin2020}. Few papers discussing phase-space behavior of the oscillator-like quantum systems with the position-dependent mass also exist~\cite{chen2006,dutra2008,cherroud2017}. In~\cite{jafarov2022a}, we computed the simplest Gaussian smoothed Wigner function for the oscillator model with a position-dependent effective mass exhibiting semiconfinement effect~\cite{jafarov2021,jafarov2022b}. That simplest definition of the computed Gaussian smoothed Wigner function of the joint quasiprobability of momentum and position also called as Husimi function is well known, too~\cite{husimi1940}. The main reason for computing the Husimi function rather than the Wigner function was also briefly covered in~\cite{jafarov2022a}. The fundamental issue was mathematical in nature. During the computation of the analytical expression of the Wigner function~\cite{wigner1932}, one observed that the integrand of its integral definition simply diverges, making further calculations impossible. However, this was not the case for the Husimi function. The Gaussian smoothing applied to the integrand of the Husimi function definition simply restricted that divergence allowing further analytical computations to be performed. Such divergences commonly appear during computations of the quantum distribution functions. For example, \cite{jafarov2008} succeeds with computation of the exact expression of the Wigner function of the one-dimensional parabose oscillator, but not its Husimi function, due to that the momentum and position operators commute in a non-canonical manner. On the other hand, \cite{jafarov2007} succeeds with the computation of the exact expression of the both Wigner and Husimi functions of the $q$-deformed harmonic oscillator.

In fact, the reality is that the phase-space description of the quantum harmonic oscillator cannot simply diverge if one applies any confinement to it. Taking this statement into account, we started to think that there is some 'lost brick' in the physics definition of the used mathematical tool and one needs to find and put that brick in its empty cell. We successfully solved this problem and are now reporting on the end result, which is the analytical expression of the Wigner function for the semiconfined quantum harmonic oscillator model under discussion.

Our paper is structured as follows. In Section 2, basic information about the Wigner function definition is briefly reviewed and then its well-known analytical expressions for a case of the nonrelativistic canonical quantum harmonic oscillator with and without the applied external homogeneous field are also presented. Section 3 is devoted to the computation of the Wigner function for the oscillator model with a position-dependent effective mass exhibiting a semiconfinement effect. These computations also are performed for both cases without and with the applied external homogeneous field. Section 4 contains detailed discussions of the obtained analytical expressions, their contour depicting, limit relations, and a brief conclusion.

\section{The Wigner joint quasiprobability distribution of the position and momentum}

As we noted in the introduction, the Wigner function plays an exceptional role in the description of any quantum system within the phase space of momentum and position, which is very similar to classical physics approaches. Its general definition for the pure stationary quantum states $\left| n \right\rangle$ in the framework of the assumption that $\hat p$ momentum and $\hat x$ position operators of the one-dimensional quantum system under consideration simply do not commute, can be written as follows~\cite{tatarskii1983}:

\be
\label{wf-nc}
W_n \left( {p,x} \right) = \frac{1}{{4\pi ^2 }}\int {\int {\left\langle n \right|e^{i\left( {\lambda \hat p + \mu \hat x} \right)} \left| n \right\rangle e^{ - i\left( {\lambda p + \mu x} \right)} d\mu d\lambda } } .
\ee

$\lambda$ and $\mu$ appearing in this definition, act as some real variables generally associated with the values of the momentum and position of the quantum system itself. Definition (\ref{wf-nc}) completely simplifies, if one takes into account the canonical commutation relation between the $\hat p$ momentum and $\hat x$ position operators of the one-dimensional quantum system, which says us that the commutation between these two operators is $\left[ {\hat p,\hat x} \right] =  - i\hbar$. Then, (\ref{wf-nc}) reduces to the well-known definition of the Wigner function empirically introduced in~\cite{wigner1932} as a method allowing to compute the quantum corrections to the thermodynamic equilibrium state of the physical system under consideration. That definition of the Wigner distribution function is the following integral consisting of the integrand of the combination of the shifted wavefunctions:

\be
\label{wf-gen}
W_n \left( {p,x} \right) = \frac{1}{{2\pi \hbar }}\int { \psi _n^* \left( {x - \frac{1}{2}x'} \right) \psi _n \left( {x + \frac{1}{2}x'} \right)e^{ - i\frac{{px'}}{\hbar }} dx'} .
\ee

Here, $ \psi _n \left( x \right)$ are orthonormalized wavefunctions of the stationary states of the quantum system under consideration in the configuration representation. A similar definition of the Wigner function can be easily written down also via the momentum representation wavefunction of the quantum system. General definition of the Wigner function (\ref{wf-nc}) or (\ref{wf-gen}) also imposes the bounded restriction $\left| {W_n \left( {p,x} \right)} \right| \le \left( {\pi \hbar } \right)^{ - 1} $ to it. Such a behavior exhibits that the function is valid for both positive and negative values of the momentum and position. Therefore, the function is called a joint quasiprobability distribution function of momentum $p$ and position $x$. However, the function defined through (\ref{wf-nc}) or (\ref{wf-gen}) is strictly positive, if the wavefunctions $\psi \left( {x} \right)$ are also strictly of the Gaussian behavior. A well-known example of such behavior is the ground state Wigner function of the non-relativistic quantum harmonic oscillator. Analytical expression of its arbitrary stationary state $n$ for this quantum system under the action of the external homogeneous field $V^{ext}\left(x \right)=gx$ can be exactly computed via its following orthonormalized wavefunctions of the stationary states

\be
\label{wf-gh}
\psi _{Nn}^g \left( x \right) = C_{Nn} e^{ - \frac{{\lambda _0 ^2 }}{2}\left( {x + x_0 } \right)^2 } H_n \left( {\lambda _0 \left( {x + x_0 } \right)} \right)\ \quad n=0,1,2,\ldots. 
\ee

Here, $H_n \left( x \right)$ is the Hermite polynomial. It is defined via the $_2F_0$ hypergeometric functions~\cite{koekoek2010}. Additionally, the following notations are introduced, too:

\[
\lambda _0  = \sqrt {\frac{{m_0 \omega }}{\hbar }} ,\quad x_0  = \frac{g}{{m_0 \omega ^2 }},\quad C_{Nn}  = \frac{{C_{N0} }}{{\sqrt {2^n n!} }},\quad C_{N0}  = \sqrt[4]{{\frac{{\lambda _0 ^2 }}{\pi }}}.
\]

Wavefunctions (\ref{wf-gh}) satisfy an orthogonality relation within the region $\left(-\infty,+\infty \right)$. Therefore, the Wigner distribution function of the non-relativistic quantum harmonic oscillator under the action of the external homogeneous field being computed via (\ref{wf-gh}) has the following analytical expression:

\be
\label{wif-gh}
W_{Nn}^g \left( {p,x} \right) = \frac{\left( { - 1} \right)^n}{{\pi \hbar}} e^{ - \frac{2}{{\hbar \omega }}\left( {\frac{{p^2 }}{{2m_0 }} + \frac{{m_0 \omega ^2 }}{2}x^2  + gx + \frac{{g^2 }}{{2m_0 \omega ^2 }}} \right)} L_n \left( {\frac{4}{{\hbar \omega }}\left( {\frac{{p^2 }}{{2m_0 }} + \frac{{m_0 \omega ^2 }}{2}x^2  + gx + \frac{{g^2 }}{{2m_0 \omega ^2 }}} \right)} \right). 
\ee

Here, $L_n \left( x \right)$ is the Laguerre polynomial defined via the $_1F_1$ hypergeometric functions~\cite{koekoek2010}. 

Analytical expression of the non-relativistic quantum harmonic oscillator Wigner distribution function without any applied external field ($g=0$) being special case of eq.(\ref{wif-gh}) is also well known~\cite{tatarskii1983}:

\be
\label{wif-h}
W_{Nn}^0 \left( {p,x} \right) = \frac{\left( { - 1} \right)^n}{{\pi \hbar}} e^{ - \frac{2}{{\hbar \omega }}\left( {\frac{{p^2 }}{{2m_0 }} + \frac{{m_0 \omega ^2 }}{2}x^2 } \right)} L_n \left( {\frac{4}{{\hbar \omega }}\left( {\frac{{p^2 }}{{2m_0 }} + \frac{{m_0 \omega ^2 }}{2}x^2 } \right)} \right).
\ee

It also can be written down via substitution of the following analytical expression of the wavefunctions of the stationary states of the non-relativistic quantum harmonic oscillator~\cite{landau1991}:

\be
\label{wf-h}
\psi _{Nn}^0 \left( x \right) = \frac{1}{{\sqrt {2^n n!} }}\left( {\frac{{\lambda _0 ^2 }}{\pi }} \right)^{\frac{1}{4}} e^{ - \frac{{\lambda _0 ^2 }}{2}x^2 } H_n \left( {\lambda _0 x} \right).
\ee

Due to that the ground state of both wavefunctions of the stationary states of the non-relativistic quantum harmonic oscillator with and without the action of the external homogeneous field (\ref{wf-gh}) and (\ref{wf-h}) are definitely of the Gaussian behavior, both Wigner functions of the ground state $W_{N0}^g \left( {p,x} \right)$ and $W_{N0}^0 \left( {p,x} \right)$ extracted from (\ref{wif-gh}) and (\ref{wif-h}) for value $n=0$  

\begin{eqnarray}
\label{wif0-gh}
W_{N0}^g \left( {p,x} \right) &=& \frac{1}{{\pi \hbar}} e^{ - \frac{2}{{\hbar \omega }}\left( {\frac{{p^2 }}{{2m_0 }} + \frac{{m_0 \omega ^2 }}{2}x^2  + gx + \frac{{g^2 }}{{2m_0 \omega ^2 }}} \right)}, \\ 
\label{wif0-h}
W_{N0}^0 \left( {p,x} \right) &=& \frac{1}{{\pi \hbar}} e^{ - \frac{2}{{\hbar \omega }}\left( {\frac{{p^2 }}{{2m_0 }} + \frac{{m_0 \omega ^2 }}{2}x^2 } \right)} ,
\end{eqnarray}
are also definitely positive.

\section{Computation of the Wigner function of a semiconfined harmonic oscillator model}

The previous section deals with the well-known one-dimensional non-relativistic canonical quantum harmonic oscillator model in the phase space. There exist a lot of interesting exact solutions of the one-dimensional quantum systems with masses depending on the position~\cite{dekar1998,carinena2004,schmidt2007,midya2010,ruby2010,levai2010,lima2012,christiansen2013,amir2014,cessa2014,ranada2014,christiansen2014,ganguly2014,ruby2015,nikitin2015,amir2016,yahiaoui2017,carinena2017,dacosta2018,halberg2018,jesus2019,amir2020,dacosta2020,dacosta2021,dacosta2023}. One of them is introduced in~\cite{jafarov2021} as an exactly-solvable model of the one-dimensional non-relativistic canonical quantum harmonic oscillator with the following effective mass varying by position:

\be
\label{mx}
M\left( x \right) = \left\{ \begin{array}{ll}
\displaystyle \frac{{am_0 }}{{a + x}}, & \hbox{ for } -a<x<+\infty \\[3mm]
+\infty, &\hbox{ for } x\leq -a
\end{array} \right.
\qquad (a>0).
\ee

That model is semiconfined, i.e. its wavefunctions of the stationary states vanish at both values of the position $x=-a$ and $x \to +\infty$, but, the energy spectrum corresponding to such behavior of the wavefunctions completely overlaps with the energy spectrum of the standard non-relativistic canonical quantum harmonic oscillator. \cite{jafarov2021} obtains the exact solution to the semiconfined harmonic oscillator model with the mass $M\left( x \right)$ varying by position in the framework of the BenDaniel–Duke kinetic energy operator  generalization~\cite{bendaniel1966}. The main feature of the BenDaniel–Duke kinetic energy operator is that being the simplest generalized version of the non-relativistic kinetic energy operator with the mass varying by position in the configuration representation, this kinetic energy operator preserves its Hermitian property. \cite{gora1969,zhu1983,vonroos1983,karthiga2017} also consider different kinetic energy operator generalizations for a case of the existence of the position-dependent mass. However, such generalizations only modify the Hamiltonian with the homogeneous parameter terms contributions if one considers harmonic oscillator potential with the varying mass behaving itself as (\ref{mx}). Despite that such a homogeneous parameter contribution can have an interest from a physics viewpoint (discontinuity of the wavefunction, restriction values of the energy, etc.), the general mathematical approach for the exact solution of the quantum harmonic oscillator system under study remains unchanged. More details of similar discussions can be found in~\cite{jafarov2021,jafarov2020}. Therefore, the exact solution to the semiconfined oscillator model with the varying mass was restricted to the case of the BenDaniel–Duke kinetic energy operator generalization. Its wavefunctions of the stationary states are expressed through the generalized Laguerre polynomials as follows:

\be
\label{wf-sc-0}
\psi _n \left( x \right) \equiv \psi _n^{SC} \left( x \right) = C_n^{SC} \left( {x + a} \right)^{\lambda _0 ^2 a^2 } e^{ - \lambda _0 ^2 a\left( {x + a} \right)} L_n^{\left( {2\lambda _0 ^2 a^2 } \right)} \left( {2\lambda _0 ^2 a\left( {x + a} \right)} \right),
\ee
where the normalization factor equals to

\be
\label{c-n}
C_n^{SC}  = \left( { - 1} \right)^n \left( {2\lambda _0 ^2 a } \right)^{\lambda _0 ^2 a^2  + \frac{1}{2}} \sqrt {\frac{{n!}}{{\Gamma \left( {n + 2\lambda _0 ^2 a^2  + 1} \right)}}} .
\ee

Next, this model also was generalized to the case of the applied external homogeneous field~\cite{jafarov2022b} and the following analytical expression of the wavefunctions of the stationary states in terms of the generalized Laguerre polynomials have been obtained:

\be
\label{wf-gsc}
\psi _n \left( x \right) \equiv \psi _n^{gSC} \left( x \right) =  C_n^{gSC} \left( {x + a} \right)^{\lambda _0 ^2 a^2 } e^{ - \lambda _0 ^2 ag_0 \left( {x + a} \right)} L_n^{\left( {2\lambda _0 ^2 a^2 } \right)} \left( {2\lambda _0 ^2 ag_0 \left( {x + a} \right)} \right),
\ee
where,

\be
\label{c-ng}
C_n^{gSC}=g_0 ^{\lambda _0 ^2 a^2  + \frac{1}{2}} C_n^{SC},
\ee
with the normalization factor $C_n^{SC}$ is same as from (\ref{c-n}) and the parameter $g_0$ is defined as

\[
g_0  = \sqrt {1 + 2\frac{{x_0}}{{ a}}} .
\]

In case of the absence of the external field corresponding to value $g=0$ ($x_0=0$), the parameter $g_0$ defined above simply equals one. Then, the wavefunction (\ref{wf-gsc}) reduces to the wavefunction (\ref{wf-sc-0}). One needs to note that both wavefunctions (\ref{wf-gsc}) and (\ref{wf-sc-0}) satisfy the following orthogonality relation:

\[
\int_{-a}^{+\infty} \psi_m(x) \psi_n(x) \,dx = \delta_{mn},
\]
which can be deduced from the known orthogonality relation of the generalized Laguerre polynomials~\cite{koekoek2010}.

Already, we computed in~\cite{jafarov2022a} the simplest realization of the Gaussian smoothing for the Wigner function for the oscillator model with a position-dependent effective mass exhibiting a semiconfinement effect~\cite{jafarov2021,jafarov2022b}. However, our initial goal was the computation of the exact expression of the Wigner function itself even without any simplest Gaussian smoothing. First of all, we took into account that the oscillator model with a position-dependent effective mass exhibiting semiconfinement effect is constructed within the non-relativistic canonical approach. Therefore, the use of the Wigner function definition (\ref{wf-gen}) instead of the more general definition (\ref{wf-nc}) was sufficient. Next, it was necessary to take into account that the wavefunctions of the stationary states (\ref{wf-sc-0}) and (\ref{wf-gsc}) vanish at both values of the position $x=-a$ and $x \to +\infty$. Therefore, the integral in the definition of the Wigner function (\ref{wf-gen}) should have integration limits from $x=-a$ to $x \to +\infty$. At that point, we observed that the integrand from eq.(\ref{wf-gen}) simply diverges and this fact makes it impossible to perform further calculations. However, Gaussian smoothing of eq.(\ref{wf-gen}) restricted that divergence and allowed us to perform the computations of the simplest Gaussian smoothed Wigner function (or Husimi function) for this semiconfined model and obtain the exact expression of the phase-space function in terms of the parabolic cylinder function~\cite{jafarov2022a}. But, actually, the phase-space description of the quantum harmonic oscillator exists from a physics viewpoint, and confinement as an effect cannot diverge it. Analytical expression of the Husimi function for the same model that we managed to compute was evidence for this statement. By thinking a little bit more deeply one could solve the divergence problem of the integrand.

We start our computation directly from the semiconfined oscillator model generalized to the case of the applied external homogeneous field and substitute its wavefunctions of the stationary states (\ref{wf-gsc}) at the definition of the Wigner function~(\ref{wf-gen}):

\begin{eqnarray}
\label{wigf-1}
 W_n^g \left( {p,x} \right) &=& \frac{{\left( {C_n^{gSC} } \right)^2 }}{{2\pi \hbar }}e^{ - 2g_0 \lambda _0 ^2 a\left( {x + a} \right)}  \\ 
  &\times& \int {e^{ - i\frac{p}{\hbar }x'} \left[ {\left( {x + a} \right)^2  - \frac{{x'^2 }}{4}} \right]^{\lambda _0 ^2 a^2 } L_n^{\left( {2\lambda _0 ^2 a^2 } \right)} \left( {2g_0 \lambda _0 ^2 a\left( {x - \frac{{x'}}{2} + a} \right)} \right)L_n^{\left( {2\lambda _0 ^2 a^2 } \right)} \left( {2g_0 \lambda _0 ^2 a\left( {x + \frac{{x'}}{2} + a} \right)} \right)dx'}. \nonumber 
\end{eqnarray}

This expression becomes more compact if one applies the change of the variable as $y=x'/2$:

\begin{eqnarray}
\label{wigf-2}
 W_n^g \left( {p,x} \right) &=& \frac{{\left( {C_n^{gSC} } \right)^2 }}{{\pi \hbar }}e^{ - 2g_0 \lambda _0 ^2 a\left( {x + a} \right)}  \\ 
  &\times& \int {e^{ - 2i\frac{p}{\hbar }y} \left[ {\left( {x + a} \right)^2  - y^2 } \right]^{\lambda _0 ^2 a^2 } L_n^{\left( {2\lambda _0 ^2 a^2 } \right)} \left( {2g_0 \lambda _0 ^2 a\left( {x + a - y} \right)} \right)L_n^{\left( {2\lambda _0 ^2 a^2 } \right)} \left( {2g_0 \lambda _0 ^2 a\left( {x + a + y} \right)} \right)dy}.  \nonumber 
\end{eqnarray}

As we noted above, if one defines the limits of the integral from $-a/2$ to $+\infty$, then the integral diverges. However, one needs to take into account that if the wavefunction $\psi _n^{gSC} \left( x \right)$ defined through expression (\ref{wf-gsc}) vanishes at finite value $x=-a$, then both $\psi _n^{gSC} \left( x+2y \right)$ and $\psi _n^{gSC} \left( x - 2y \right)$ forming the integrand of the Wigner function also have to vanish at finite value $x=-a$. This is a main difference of the Wigner function from any of its Gaussian smoothed analogues, which does not exhibit itself if one deals with the phase space of the quantum system defined within the whole region $\left(-\infty,+\infty\right)$ (i.e. the quantum system with the wavefunctions vanishing at $\pm \infty$ values of the position and momentum). Taking this property into account one obtains that the integral limits are within $- \left( {x + a} \right) \le y \le x + a$. Therefore, the modified version of the Wigner function (\ref{wigf-2}) is as follows:

\begin{eqnarray}
\label{wigf-3}
 W_n^g \left( {p,x} \right) &=& \frac{{\left( {C_n^{gSC} } \right)^2 }}{{\pi \hbar }}e^{ - 2g_0 \lambda _0 ^2 a\left( {x + a} \right)}  \\ 
  &\times& \int\limits_{ - \left( {x + a} \right)}^{x + a} {e^{ - 2i\frac{p}{\hbar }y} \left[ {\left( {x + a} \right)^2  - y^2 } \right]^{\lambda _0 ^2 a^2 } L_n^{\left( {2\lambda _0 ^2 a^2 } \right)} \left( {2g_0 \lambda _0 ^2 a\left( {x + a - y} \right)} \right)L_n^{\left( {2\lambda _0 ^2 a^2 } \right)} \left( {2g_0 \lambda _0 ^2 a\left( {x + a + y} \right)} \right)dy}.  \nonumber 
\end{eqnarray}

Now, the integrand does not diverge under these integration limits and the above integral is analytically computable. Further, we slightly change the variable $y$ to $t$ as follows:

\[
t = \frac{y}{{x + a}}.
\] 

Its substitution at (\ref{wigf-3}) yields:

\begin{eqnarray}
\label{wigf-4}
 W_n^g \left( {p,x} \right) &=& \frac{{\left( {C_n^{gSC} } \right)^2 }}{{\pi \hbar }}\left( {x + a} \right)^{2\lambda _0 ^2 a^2  + 1} e^{ - 2g_0 \lambda _0 ^2 a\left( {x + a} \right)}  \\ 
  &\times& \int\limits_{ - 1}^1 {e^{ - 2i\frac{p}{\hbar }\left( {x + a} \right)t} \left( {1 - t^2 } \right)^{\lambda _0 ^2 a^2 } L_n^{\left( {2\lambda _0 ^2 a^2 } \right)} \left( {2g_0 \lambda _0 ^2 a\left( {x + a} \right)\left( {1 - t} \right)} \right)L_n^{\left( {2\lambda _0 ^2 a^2 } \right)} \left( {2g_0 \lambda _0 ^2 a\left( {x + a} \right)\left( {1 + t} \right)} \right)dt} . \nonumber
\end{eqnarray}

First of all, it is convenient to analyze the ground state distribution function. Therefore, one considers the value $n=0$. Then, eq.(\ref{wigf-4}) simplifies as follows:

\be
\label{wigfg0-1}
W_0^g \left( {p,x} \right) = \frac{{\left( {C_0^{gSC} } \right)^2 }}{{\pi \hbar }}\left( {x + a} \right)^{2\lambda _0 ^2 a^2  + 1} e^{ - 2g_0 \lambda _0 ^2 a\left( {x + a} \right)} \int\limits_{ - 1}^1 {e^{ - 2i\frac{p}{\hbar }\left( {x + a} \right)t} \left( {1 - t^2 } \right)^{\lambda _0 ^2 a^2 } dt} .
\ee

Integral appearing here can be computed further exactly. For this reason, one needs first to replace the exponential function from the integrand with its following Maclaurin expansion:

\[
e^{ - 2i\frac{p}{\hbar }\left( {x + a} \right)t}  = \sum\limits_{k = 0}^\infty  {\frac{{\left( { - 2i\frac{p}{\hbar }\left( {x + a} \right)} \right)^k }}{{k!}}t^k } .
\]

One obtains that

\be
\label{wigfg0-11}
W_0^g \left( {p,x} \right) = \frac{{\left( {C_0^{gSC} } \right)^2 }}{{\pi \hbar }}\left( {x + a} \right)^{2\lambda _0 ^2 a^2  + 1} e^{ - 2g_0 \lambda _0 ^2 a\left( {x + a} \right)} \sum\limits_{k = 0}^\infty  {\frac{{\left( { - 2i\frac{p}{\hbar }\left( {x + a} \right)} \right)^k }}{{k!}}\int\limits_{ - 1}^1 {\left( {1 - t^2 } \right)^{\lambda _0 ^2 a^2 } t^k dt} } .
\ee

Integral appearing in (\ref{wigfg0-11}) can be exactly computed in terms of the Gamma functions yielding:

\[
\int\limits_{ - 1}^1 {\left( {1 - t^2 } \right)^{\lambda _0 ^2 a^2 } t^k dt}  = \frac{1}{2}\left( {1 + \left( { - 1} \right)^k } \right)\frac{{\Gamma \left( {\frac{{k + 1}}{2}} \right)\Gamma \left( {\lambda _0 ^2 a^2  + 1} \right)}}{{\Gamma \left( {\lambda _0 ^2 a^2  + \frac{{k + 3}}{2}} \right)}}.
\]

Its substitution at (\ref{wigfg0-11}) leads to a reduction of the odd terms of the expansion over $k$ due to multiplication to $\left( {1 + \left( { - 1} \right)^k } \right)$. Then, such a reduction simplifies the ground state Wigner distribution function as follows:

\be
\label{wigfg0-12}
W_0^g \left( {p,x} \right) = \frac{{\left( {C_0^{gSC} } \right)^2 }}{{\pi \hbar }} \Gamma \left( {\lambda _0 ^2 a^2  + 1} \right) \left( {x + a} \right)^{2\lambda _0 ^2 a^2  + 1} e^{ - 2g_0 \lambda _0 ^2 a\left( {x + a} \right)} \sum\limits_{m = 0}^\infty  {\frac{{\left( { - 2i\frac{p}{\hbar }\left( {x + a} \right)} \right)^{2m} }}{{\left( {2m} \right)!}}\frac{{\Gamma \left( {m + \frac{1}{2}} \right)}}{{\Gamma \left( {\lambda _0 ^2 a^2  + m + \frac{3}{2}} \right)}}} .
\ee

The ratio of two Gamma functions appearing in the above expansions can be reexpressed as follows:

\[
\frac{{\Gamma \left( {m + \frac{1}{2}} \right)}}{{\Gamma \left( {\lambda _0 ^2 a^2  + m + \frac{3}{2}} \right)}} = \frac{{\sqrt \pi  }}{{\Gamma \left( {\lambda _0 ^2 a^2  + \frac{3}{2}} \right)}}\frac{{\left( {\frac{1}{2}} \right)_m }}{{\left( {\lambda _0 ^2 a^2  + \frac{3}{2}} \right)_m }}.
\]

Its substitution at (\ref{wigfg0-12}) yields:

\be
\label{wigfg0-13}
W_0^g \left( {p,x} \right) = \frac{{\left( {C_0^{gSC} } \right)^2 }}{{\sqrt \pi \hbar }} \frac{{\Gamma \left( {\lambda _0 ^2 a^2  + 1} \right)}}{{\Gamma \left( {\lambda _0 ^2 a^2  + \frac{3}{2}} \right)}} \left( {x + a} \right)^{2\lambda _0 ^2 a^2  + 1} e^{ - 2g_0 \lambda _0 ^2 a\left( {x + a} \right)} \sum\limits_{m = 0}^\infty  {\frac{{\left( {\frac{1}{2}} \right)_m }}{{\left( {\lambda _0 ^2 a^2  + \frac{3}{2}} \right)_m }}\frac{{\left( { - 2i\frac{p}{\hbar }\left( {x + a} \right)} \right)^{2m} }}{{\left( {2m} \right)!}}} .
\ee

Now, taking into account that
\[
\left( {C_0^{gSC} } \right)^2  = \frac{{\left( {2g_0 \lambda _0 ^2 a} \right)^{2\lambda _0 ^2 a^2  + 1} }}{{\Gamma \left( {2\lambda _0 ^2 a^2  + 1} \right)}},
\]
one needs to apply here the following well-known relation for the Gamma functions

\[
\Gamma \left( {2z} \right) = \frac{{2^{2z - 1} }}{{\sqrt \pi  }}\Gamma \left( z \right)\Gamma \left( {z + \frac{1}{2}} \right),
\]
and even number factorials:

\[
\left( {2m} \right)! = 2^{2m} \left( {\frac{1}{2}} \right)_m m!.
\]

As a result of their use, one obtains the following analytical expression for the ground state Wigner distribution function in terms of $_0F_1$ hypergeometric function:

\be
\label{wigfg0-14}
W_0^g \left( {p,x} \right) = \frac{2}{\hbar }\frac{{\left( {g_0 \lambda _0 ^2 a} \right)^{2\lambda _0 ^2 a^2  + 1} }}{{\Gamma \left( {\lambda _0 ^2 a^2  + \frac{3}{2}} \right)\Gamma \left( {\lambda _0 ^2 a^2  + \frac{1}{2}} \right)}}\left( {x + a} \right)^{2\lambda _0 ^2 a^2  + 1} e^{ - 2g_0 \lambda _0 ^2 a\left( {x + a} \right)} \,_0 F_1 \left( {\begin{array}{*{20}c}
    -   \\
   {\lambda _0 ^2 a^2  + \frac{3}{2}}  \\
\end{array}; - \frac{{p^2 }}{{\hbar ^2 }}\left( {x + a} \right)^2 } \right).
\ee

Now, taking into account that the following analytical expression for the Bessel functions of the first kind exists:

\[
J_\alpha  \left( z \right) = \frac{{\left( {\frac{z}{2}} \right)^\alpha  }}{{\Gamma \left( {\alpha  + 1} \right)}}\,_0 F_1 \left( {\begin{array}{*{20}c}
    -   \\
   {\alpha  + 1}  \\
\end{array}; - \frac{{z^2 }}{4}} \right),
\]
one obtains that

\be
\label{wigfg0-2}
W_0^g \left( {p,x} \right) = 2\hbar ^{\lambda _0 ^2 a^2  - \frac{1}{2}} \frac{{\left( {g_0 \lambda _0 ^2 a} \right)^{2\lambda _0 ^2 a^2  + 1} }}{{\Gamma \left( {\lambda _0 ^2 a^2  + \frac{1}{2}} \right)}}e^{ - 2g_0 \lambda _0 ^2 a\left( {x + a} \right)} \left( {\frac{{x + a}}{p}} \right)^{\lambda _0 ^2 a^2  + \frac{1}{2}} J_{\lambda _0 ^2 a^2  + \frac{1}{2}} \left( {2\frac{p}{\hbar }\left( {x + a} \right)} \right).
\ee

This result also can be obtained via direct use of the following known table integral~\cite[eq.(2.3.5.3)]{prudnikov2002-1} in terms of the Gamma function $\Gamma \left( \beta  \right)$ and Bessel function of the first kind $J_{\alpha} \left( z \right)$:

\be
\label{pbm1}
\int\limits_{ - a}^a {\left( {a^2  - x^2 } \right)^{\beta  - 1} e^{i\lambda x} dx}  = \sqrt \pi  \Gamma \left( \beta  \right)\left( {\frac{{2a}}{\lambda }} \right)^{\beta  - \frac{1}{2}} J_{\beta  - \frac{1}{2}} \left( {a\lambda } \right), \quad a>0, \; \Re \left(\beta \right) >0.
\ee

It is interesting to note that the external field does not exhibit itself through the Bessel function of the first kind. This is due to that, it does not exist anymore in the integrand. Therefore, for the case of the absence of the external field, the parameter $g_0$ becomes zero, and the Wigner function of the ground state (\ref{wigfg0-2}) slightly simplifies as follows:

\be
\label{wigf00-2}
W_0^0 \left( {p,x} \right) = 2\hbar ^{\lambda _0 ^2 a^2  - \frac{1}{2}} \frac{{\left( {\lambda _0 ^2 a} \right)^{2\lambda _0 ^2 a^2  + 1} }}{{\Gamma \left( {\lambda _0 ^2 a^2  + \frac{1}{2}} \right)}}e^{ - 2 \lambda _0 ^2 a\left( {x + a} \right)} \left( {\frac{{x + a}}{p}} \right)^{\lambda _0 ^2 a^2  + \frac{1}{2}} J_{\lambda _0 ^2 a^2  + \frac{1}{2}} \left( {2\frac{p}{\hbar }\left( {x + a} \right)} \right).
\ee

Taking into account that the Wigner function of the ground state is exactly computed in terms of the Bessel function of the first kind, then one can try to compute its analytical expression for arbitrarily excited states $n$. Therefore, one needs to go back to the expression (\ref{wigf-4}). Its integrand mainly consists of the product of two Laguerre polynomials with different arguments. One applies there the following known finite sum for such kind of products~\cite{bailey1936}:

\be
\label{pr-lp}
L_n^{\left( \alpha  \right)} \left( x \right)L_n^{\left( \alpha  \right)} \left( y \right) = \frac{{\Gamma \left( {n + \alpha  + 1} \right)}}{{n!}}\sum\limits_{k = 0}^n {\frac{{\left( {xy} \right)^k }}{{k!\Gamma \left( {k + \alpha  + 1} \right)}}L_{n - k}^{\left( {\alpha  + 2k} \right)} \left( {x + y} \right)}.
\ee

Its substitution at (\ref{wigf-4}) yields:

\begin{eqnarray}
\label{wigf-5}
 W_n^g \left( {p,x} \right) &=& \frac{{\left( {C_n^{gSC} } \right)^2 }}{{\pi \hbar }}e^{ - 2g_0 \lambda _0 ^2 a\left( {x + a} \right)} \left( {x + a} \right)^{2\lambda _0 ^2 a^2  + 1} \frac{{\Gamma \left( {n + 2\lambda _0 ^2 a^2  + 1} \right)}}{{n!}}  \\ 
 &\times& \int\limits_{ - 1}^1 {e^{ - 2i\frac{p}{\hbar }\left( {x + a} \right)t} \left( {1 - t^2 } \right)^{\lambda _0 ^2 a^2 } \sum\limits_{k = 0}^n {\frac{{\left( {2g_0 \lambda _0 ^2 a\left( {x + a} \right)} \right)^{2k} }}{{k!\Gamma \left( {k + 2\lambda _0 ^2 a^2  + 1} \right)}}\left( {1 - t^2 } \right)^k L_{n - k}^{\left( {2\lambda _0 ^2 a^2  + 2k} \right)} \left( {4g_0 \lambda _0 ^2 a\left( {x + a} \right)} \right)} dt} . \nonumber
\end{eqnarray}

Next, one interchanges integral and finite summation, which also changes eq.(\ref{wigf-5}) as follows:

\begin{eqnarray}
\label{wigf-6}
 W_n^g \left( {p,x} \right) &=& \frac{{\left( {C_n^{gSC} } \right)^2 }}{{\pi \hbar }}e^{ - 2g_0 \lambda _0 ^2 a\left( {x + a} \right)} \left( {x + a} \right)^{2\lambda _0 ^2 a^2  + 1} \frac{{\Gamma \left( {n + 2\lambda _0 ^2 a^2  + 1} \right)}}{{n!}}  \\ 
&\times& \sum\limits_{k = 0}^n {\frac{{\left( {2g_0 \lambda _0 ^2 a\left( {x + a} \right)} \right)^{2k} }}{{k!\Gamma \left( {k + 2\lambda _0 ^2 a^2  + 1} \right)}}L_{n - k}^{\left( {2\lambda _0 ^2 a^2  + 2k} \right)} \left( {4g_0 \lambda _0 ^2 a\left( {x + a} \right)} \right)\int\limits_{ - 1}^1 {e^{ - 2i\frac{p}{\hbar }\left( {x + a} \right)t} \left( {1 - t^2 } \right)^{\lambda _0 ^2 a^2  + k} dt} } . \nonumber
\end{eqnarray}

Finally, one can again successfully apply the table integral~(\ref{pbm1}) that yields an exact expression for the Wigner function of the arbitrary semiconfined quantum harmonic oscillator stationary states in the presence of the homogeneous external field:

\begin{eqnarray}
\label{wigf-7}
 W_n^g \left( {p,x} \right) &=& 2\hbar ^{\lambda _0 ^2 a^2  - \frac{1}{2}} \frac{{\left( {g_0 \lambda _0 ^2 a} \right)^{2\lambda _0 ^2 a^2  + 1} }}{{\Gamma \left( {\lambda _0 ^2 a^2  + \frac{1}{2}} \right)}}e^{ - 2g_0 \lambda _0 ^2 a\left( {x + a} \right)} \left( {\frac{{x + a}}{p}} \right)^{\lambda _0 ^2 a^2  + \frac{1}{2}}  \\ 
  &\times& \sum\limits_{k = 0}^n {\frac{{\left( {2g_0 \lambda _0 ^2 a} \right)^{2k} }}{{k!}}\frac{{\left( {\lambda _0 ^2 a^2  + 1} \right)_k }}{{\left( {2\lambda _0 ^2 a^2  + 1} \right)_k }}\left( {\hbar \frac{{x + a}}{p}} \right)^k J_{\lambda _0 ^2 a^2  + k + \frac{1}{2}} \left( {2\frac{p}{\hbar }\left( {x + a} \right)} \right)L_{n - k}^{\left( {2\lambda _0 ^2 a^2  + 2k} \right)} \left( {4g_0 \lambda _0 ^2 a\left( {x + a} \right)} \right)} . \nonumber
	\end{eqnarray}

Absence of the external field again slightly simplifies (\ref{wigf-7}) due to that $g=0$ ($g_0=1$):

\begin{eqnarray}
\label{wigf-8}
 W_n^0 \left( {p,x} \right) &=& 2\hbar ^{\lambda _0 ^2 a^2  - \frac{1}{2}} \frac{{\left( {\lambda _0 ^2 a} \right)^{2\lambda _0 ^2 a^2  + 1} }}{{\Gamma \left( {\lambda _0 ^2 a^2  + \frac{1}{2}} \right)}}e^{ - 2\lambda _0 ^2 a\left( {x + a} \right)} \left( {\frac{{x + a}}{p}} \right)^{\lambda _0 ^2 a^2  + \frac{1}{2}}  \\ 
  &\times& \sum\limits_{k = 0}^n {\frac{{\left( {2\lambda _0 ^2 a} \right)^{2k} }}{{k!}}\frac{{\left( {\lambda _0 ^2 a^2  + 1} \right)_k }}{{\left( {2\lambda _0 ^2 a^2  + 1} \right)_k }}\left( {\hbar \frac{{x + a}}{p}} \right)^k J_{\lambda _0 ^2 a^2  + k + \frac{1}{2}} \left( {2\frac{p}{\hbar }\left( {x + a} \right)} \right)L_{n - k}^{\left( {2\lambda _0 ^2 a^2  + 2k} \right)} \left( {4\lambda _0 ^2 a\left( {x + a} \right)} \right)} . \nonumber
\end{eqnarray}

We obtained an exact expression of the Wigner function of the semiconfined quantum harmonic oscillator under the action of the external homogeneous field. In the next section, its main properties as well as the behavior of the semiconfined quantum harmonic oscillator model in the phase space will be briefly discussed.

\section{Discussions}

\begin{figure}[t!]
\begin{center}
\resizebox{0.32\textwidth}{!}{%
  \includegraphics{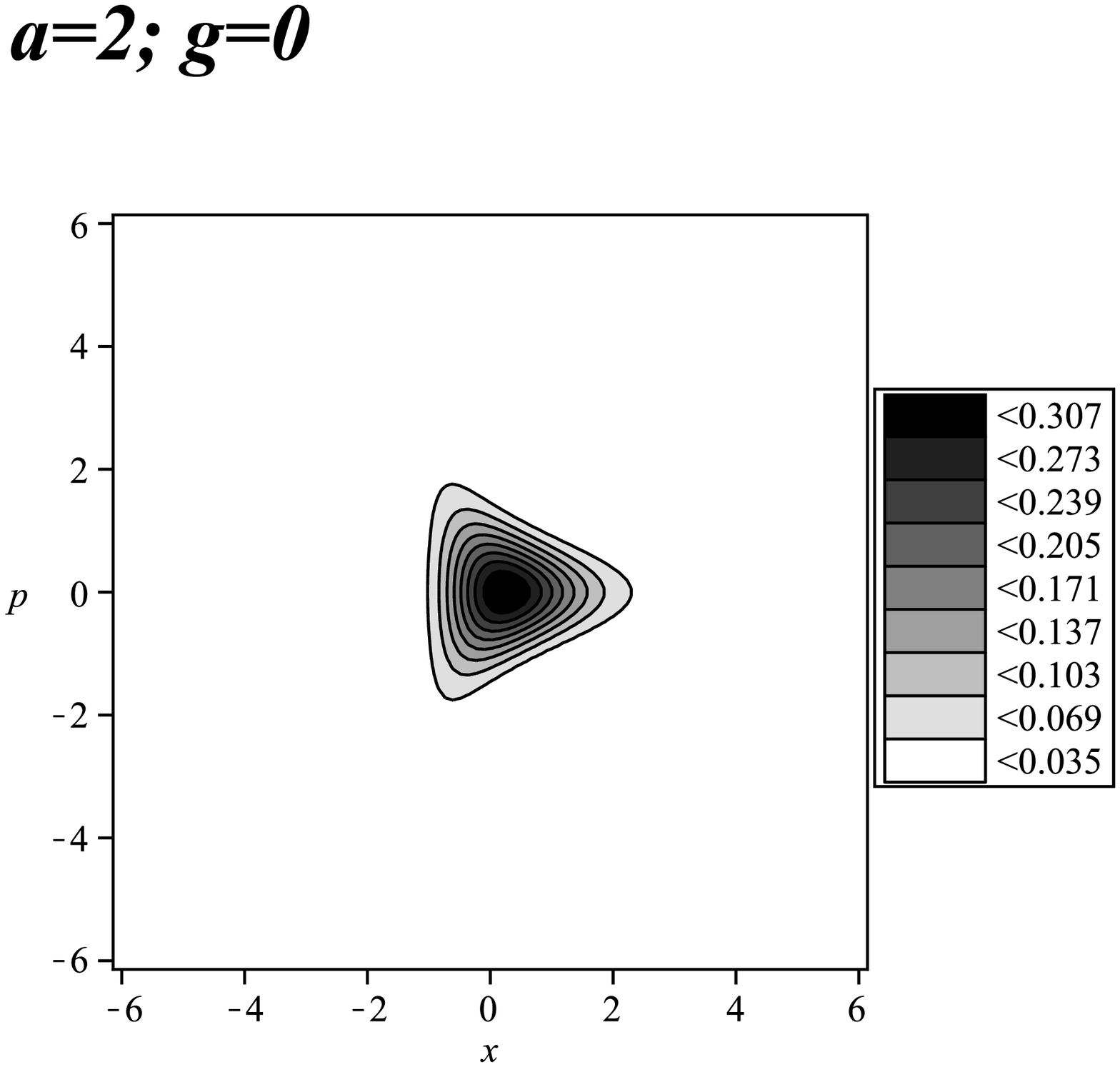}
}
\resizebox{0.32\textwidth}{!}{%
  \includegraphics{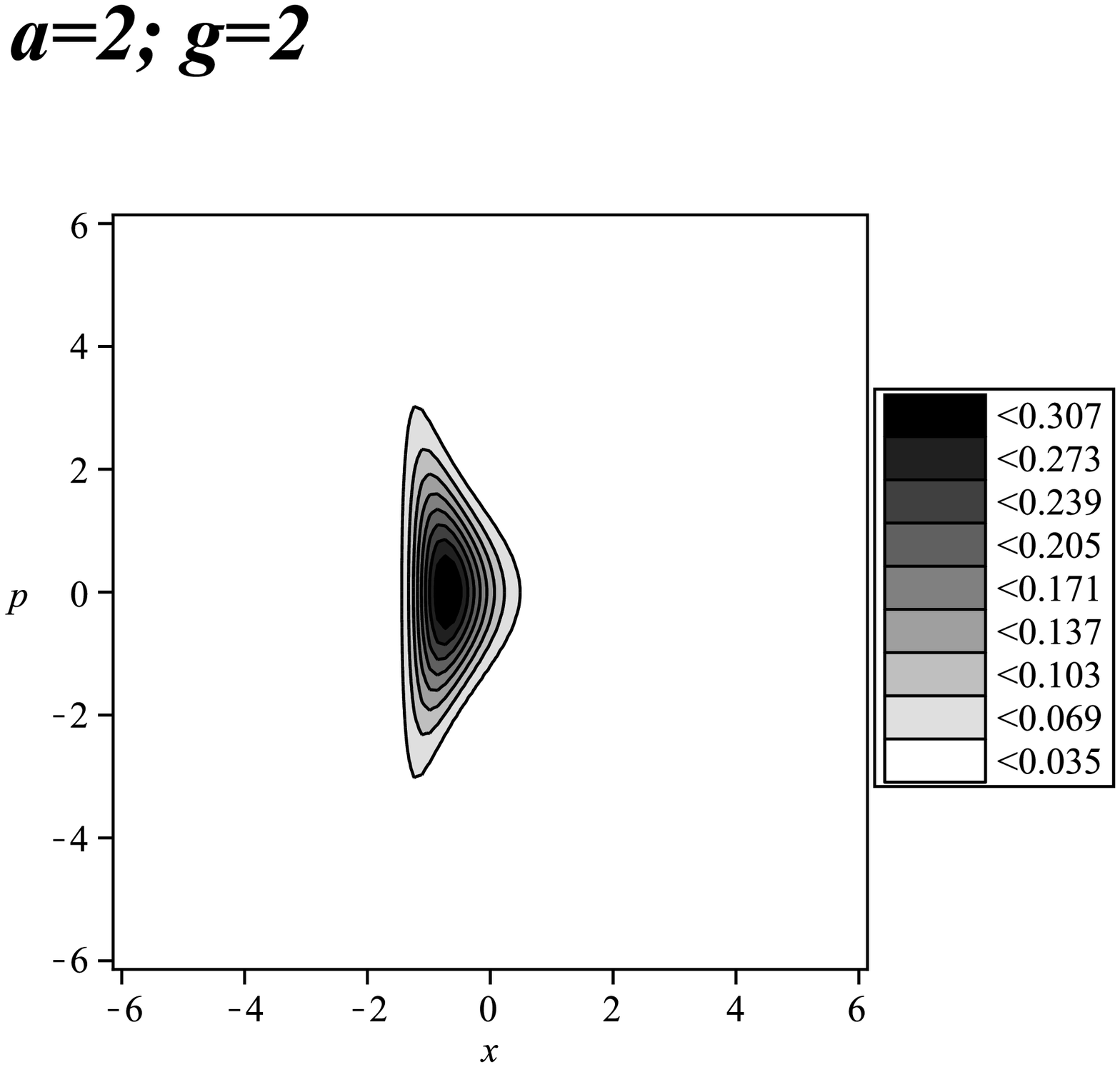}
}
\resizebox{0.32\textwidth}{!}{%
  \includegraphics{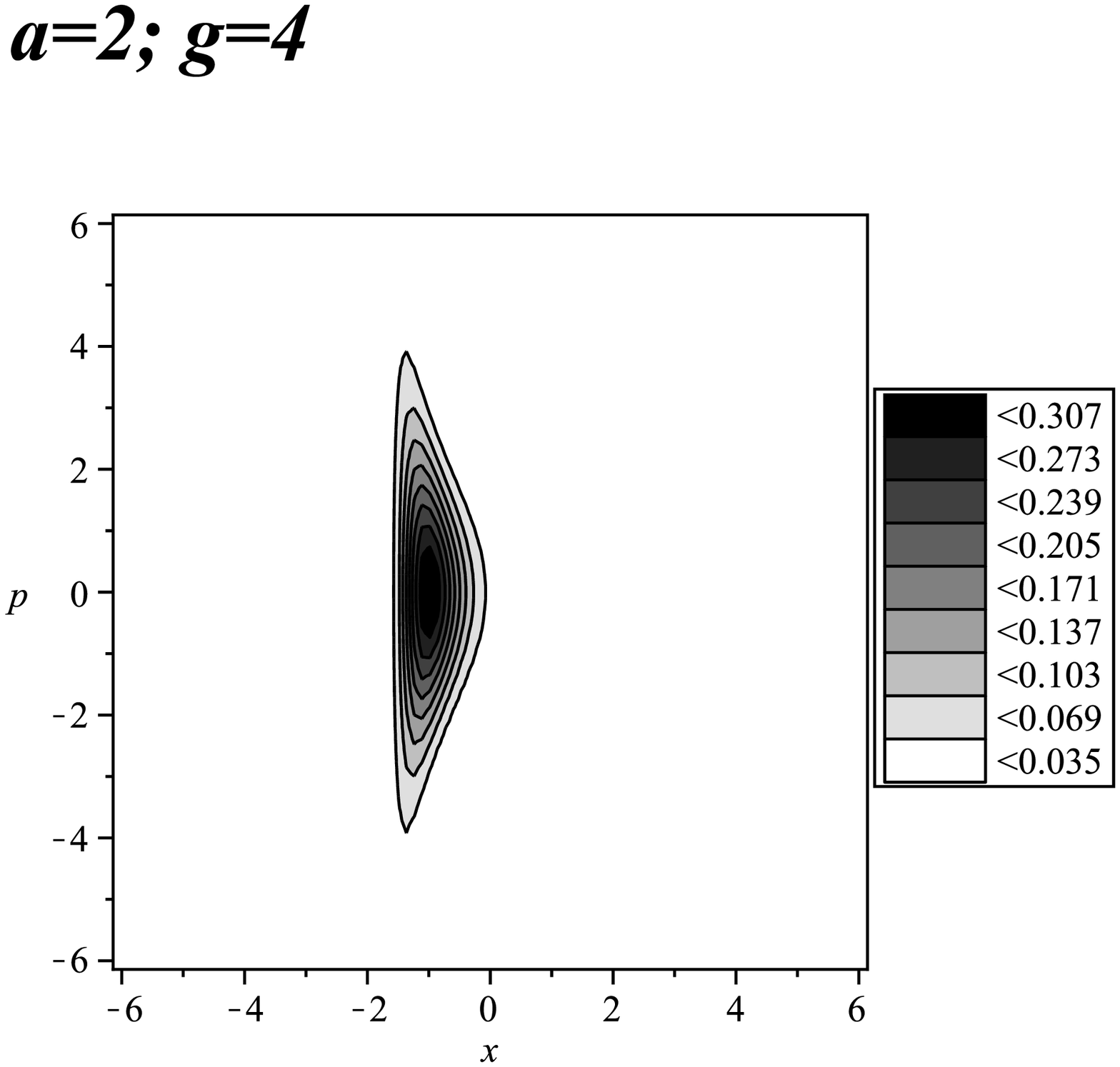}
}\\
\resizebox{0.32\textwidth}{!}{%
  \includegraphics{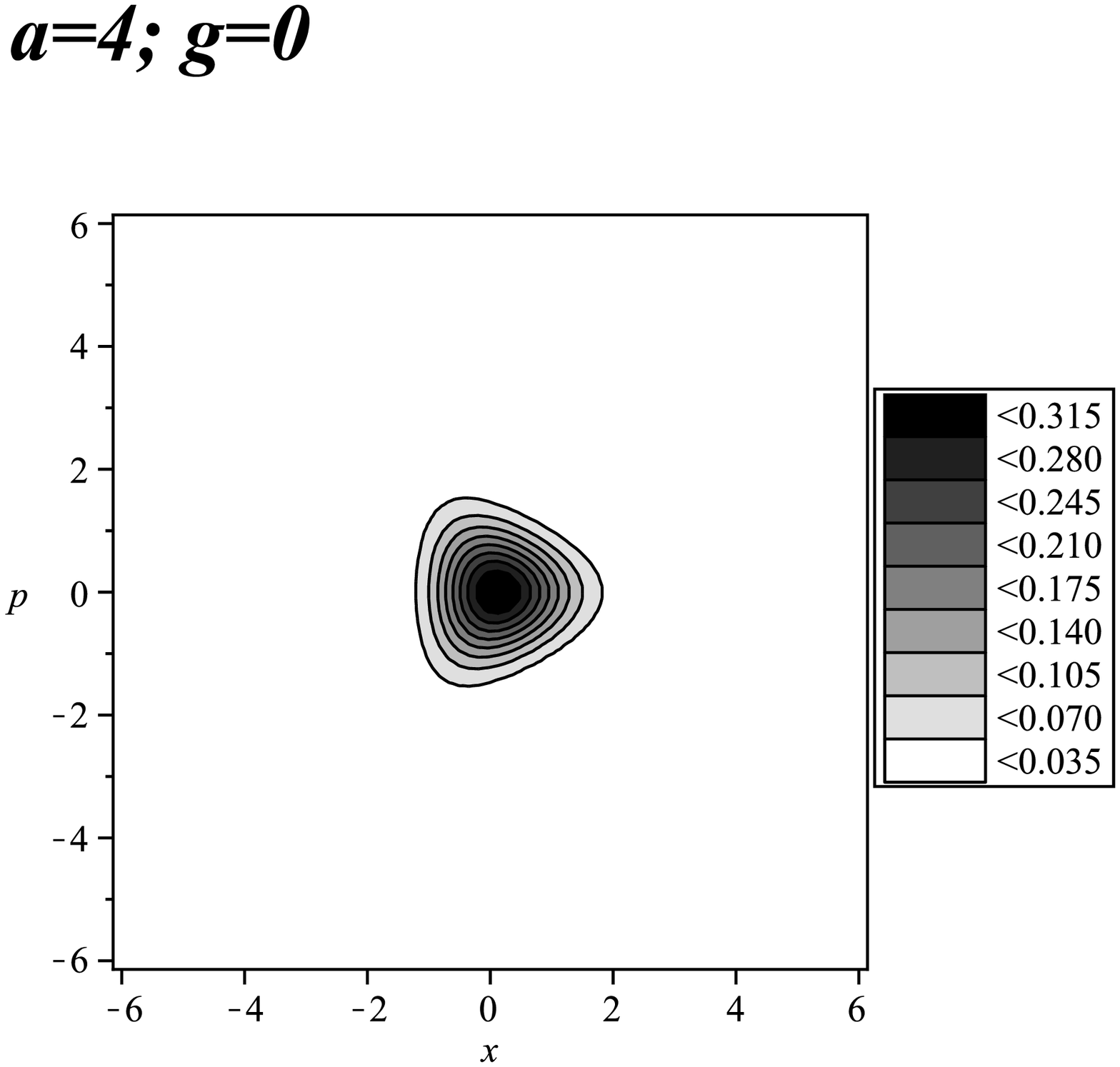}
}
\resizebox{0.32\textwidth}{!}{%
  \includegraphics{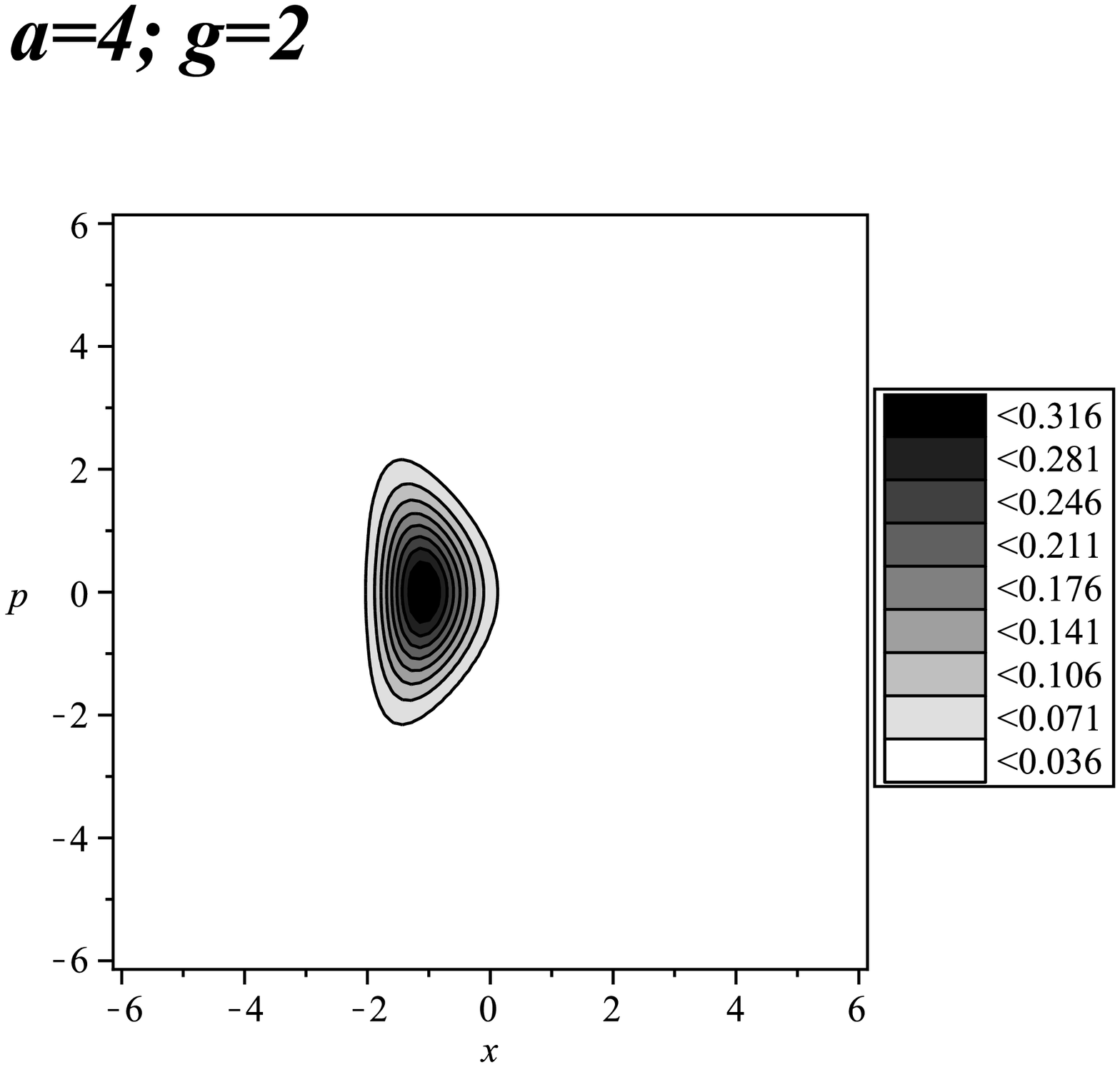}
}
\resizebox{0.32\textwidth}{!}{%
  \includegraphics{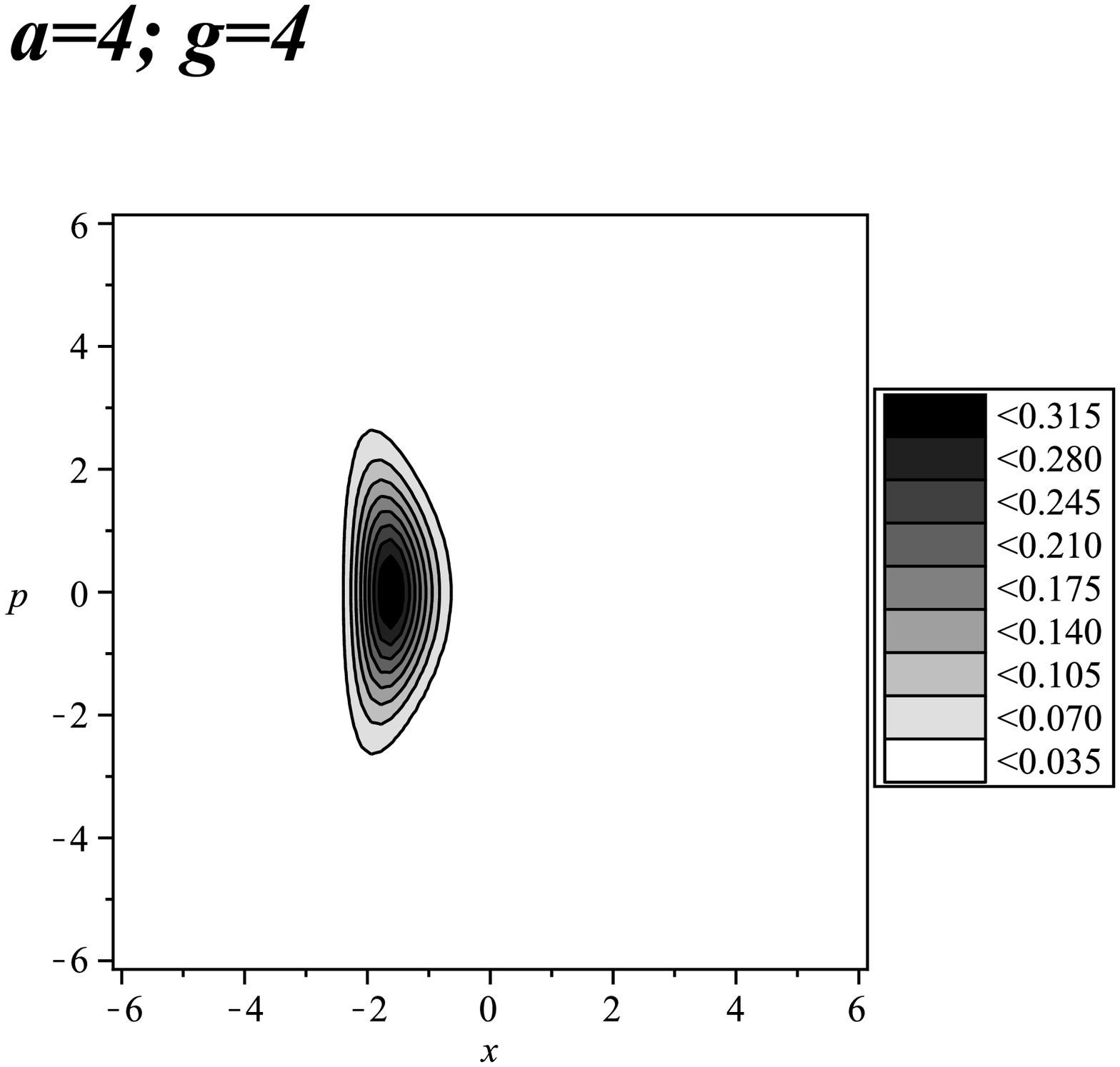}
}\\
\resizebox{0.32\textwidth}{!}{%
  \includegraphics{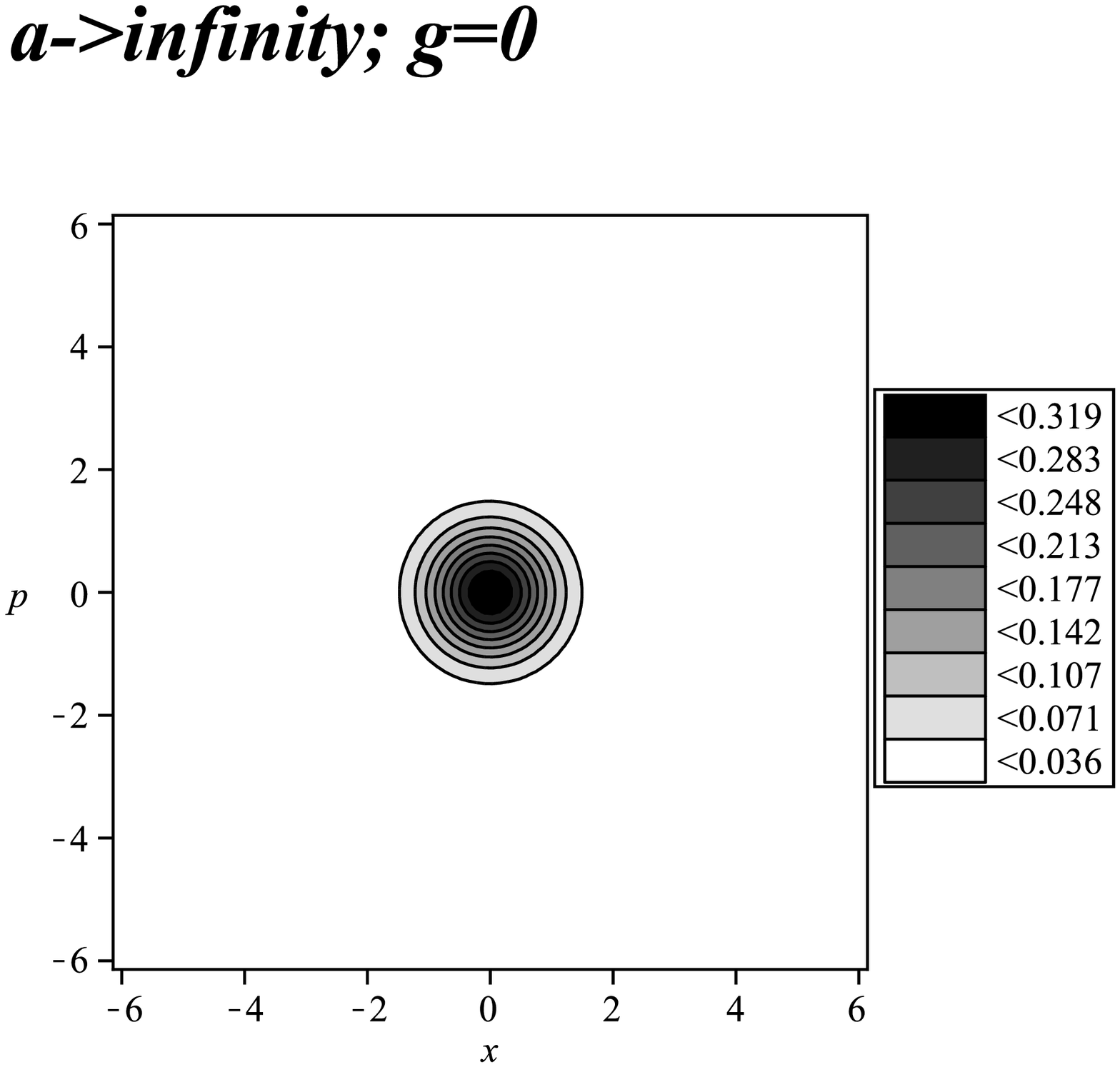}
}
\resizebox{0.32\textwidth}{!}{%
  \includegraphics{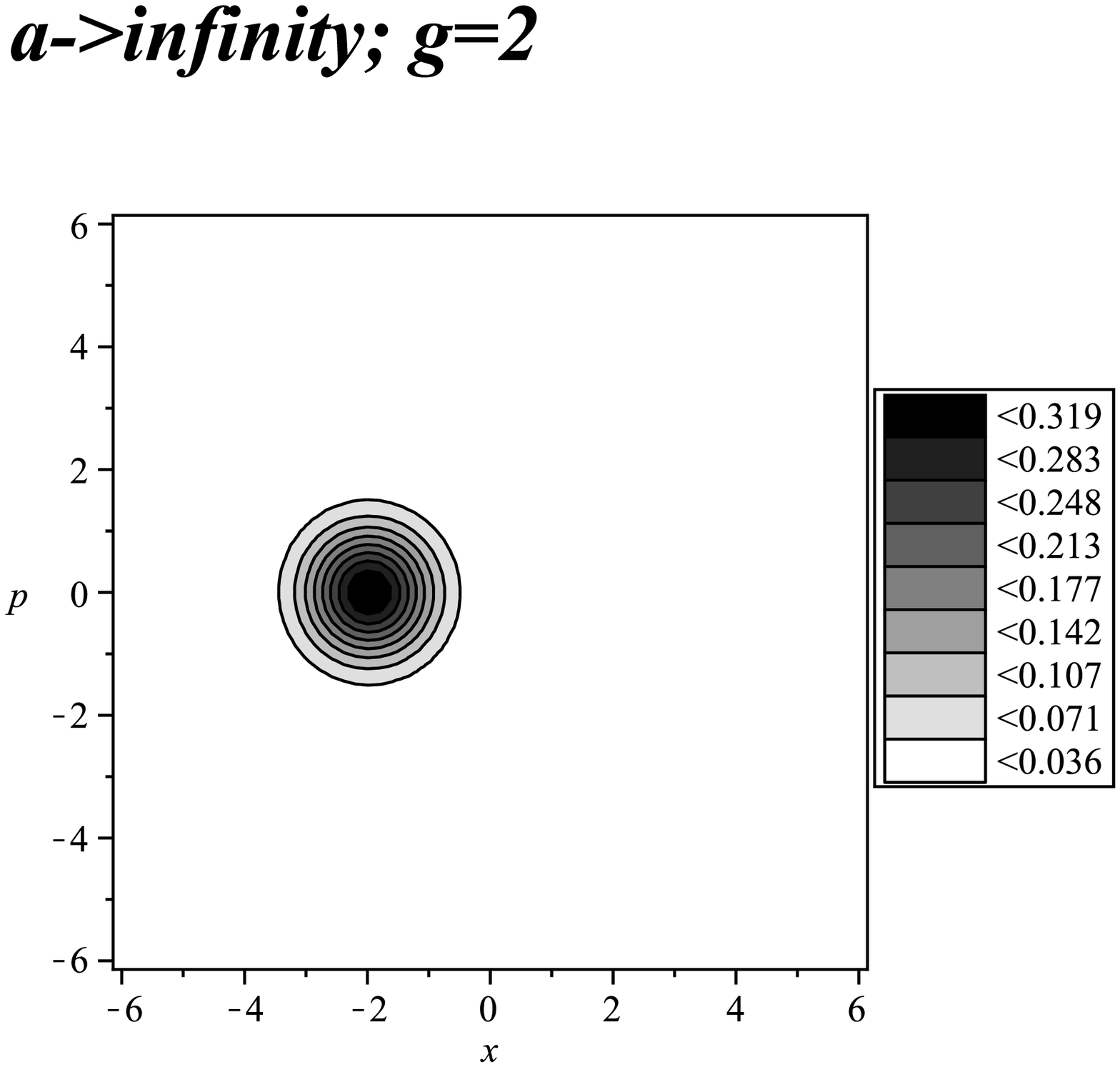}
}
\resizebox{0.32\textwidth}{!}{%
  \includegraphics{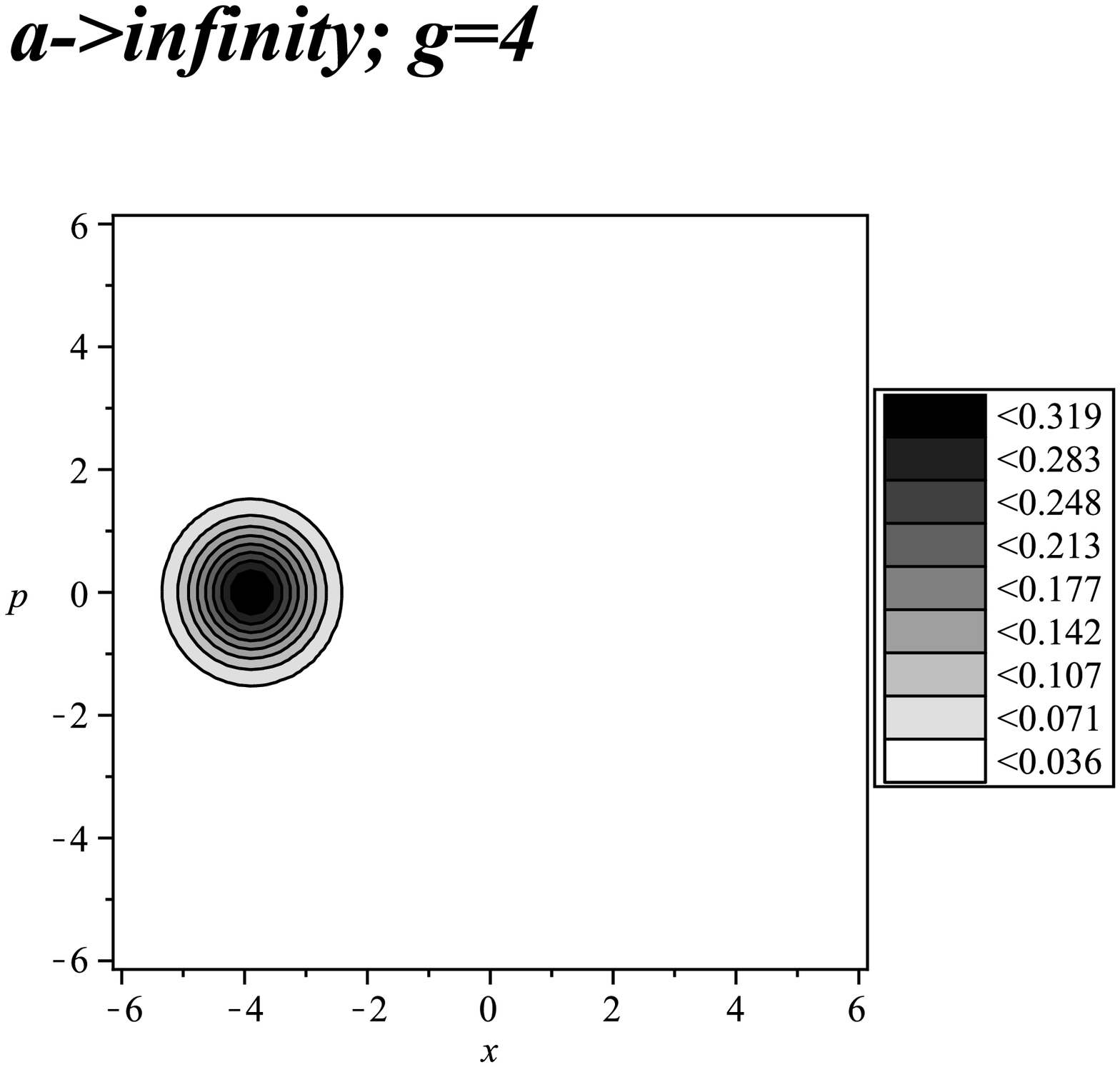}
}
\end{center}
\caption{A comparative plot of the semiconfined quantum harmonic oscillator Wigner function~(\ref{wigfg0-2}) of the ground state ($n=0$) without an external field ($g=0$, left plots) and with an external field ($g=2,\;4$; middle and right plots). Upper plots correspond to the confinement parameter $a=2$, whereas, middle plots correspond to the confinement parameter $a=4$ and lower plots correspond to the confinement parameter $a \to \infty$ ($m_0=\omega=\hbar=1$).} 
\label{fig.1}
\end{figure}

In the previous section, an exact analytical expression of the Wigner quasiprobability distribution function of the semiconfined quantum harmonic oscillator was computed through the wavefunctions of the stationary states (\ref{wf-gsc}) of this oscillator model. We considered two cases: the existence of the external homogeneous field and its absence (the special case, when $g=0$). As a result, we obtained two expressions of the Wigner function of the arbitrary $n$ states: they are the Wigner quasiprobability distribution function under the action of the external field (\ref{wigf-7}) and in the case of the absence of such a field (\ref{wigf-8}). The best tool for a deeper understanding of their behavior and possibly 'hidden' differences from the canonical harmonic oscillator Wigner function analytical expressions (\ref{wif-gh}) and (\ref{wif-h}) is a graphical visualization of these analytical expressions.

In fig.1, we decided to restrict depictions with the ground state Wigner functions (\ref{wigfg0-2}) and (\ref{wigf00-2}). These functions are definitely positive and being simpler cases of the Wigner functions of the arbitrary $n$ states (\ref{wigf-7}) and (\ref{wigf-8}), they are sufficient for qualitative analysis of the phase space features of the semiconfined oscillator model. Nine plots are presented, where dependencies of the ground state Wigner function from the parameters $a$ and $g$ are depicted. Values of $a=2;4$ and $a\to \infty$ as well as $g=0;2;4$ are considered. In the case of $a=2$, one observes squeezed Gaussian distribution around the equilibrium position-momentum value due to that position being semiconfined, but it is not valid for momentum, too. As a result, the semiconfinement of the position values causes the position distribution non-symmetrical, but it does not affect to the symmetrical distribution behavior of the momentum, which can be seen from the upper left plot. When an external field is applied to the semiconfined quantum system under study (as seen in the upper middle and right figures), its Wigner function behaves radically different than in the case of the canonical harmonic oscillator. This occurs due to that there exists an infinitely high wall at some negative position value. It simply reduces positive position values distribution, and preserves symmetric behavior, but extends the probability to both positive and negative values of the momentum. Greater values of semiconfinement parameter $a$ simply recover the symmetric Gaussian behavior of the position in the phase space. Such a behavior is a weak signature of the correct extension of the non-relativistic canonical harmonic oscillator phase space in terms of its Wigner function. Finally, the value $a \to \infty$ corresponding to lower plots completely recovers the non-relativistic canonical harmonic oscillator, in which the Wigner function is an analytical expression (\ref{wif-gh}).

As one observes from fig.1, there should be a correct limit from (\ref{wigfg0-2}) to (\ref{wif-gh}). We have to use the Stirling approximation of the Gamma function

\[
\Gamma \left( {z + 1} \right) \cong \sqrt {2\pi z} e^{z\ln z - z} ,\quad \left| z \right| \to \infty,
\]
the following asymptotics of the Bessel function of the first kind

\[
J_\alpha  \left( z \right) = \frac{1}{{\pi \sqrt 2 \sqrt[4]{{\alpha ^2  - z^2 }}}}\exp \left( {\sqrt {\alpha ^2  - z^2 }  - \alpha \arccosh \frac{\alpha }{z}} \right),\;\alpha  > z > 0,\quad \alpha \to \infty,
\]
and the following expansions ($z <  < 1$):

\begin{eqnarray}
 \frac{1}{{z + 1}} &\approx& 1 - z + z^2  -  \cdots ,\nonumber \\ 
 \frac{1}{{\sqrt {z + 1} }} &\approx& 1 - \frac{z}{2} + \frac{3}{8}z^2  -  \cdots ,\nonumber \\ 
 \sqrt {z + 1}  &\approx& 1 + \frac{z}{2} - \frac{{z^2 }}{8} +  \cdots , \nonumber\\ 
 \ln \left( {z + 1} \right) &\approx& z - \frac{{z^2 }}{2} +  -  \cdots . \nonumber 
\end{eqnarray}

Substitution of above-listed approximations, expansions, and asymptotics at (\ref{wigfg0-2}) and further straightforward long computations will lead to the complete recovery of the Wigner function (\ref{wif-gh}) under the limit $a \to \infty$.

Let's prove the correctness of the above-noted limit recovery in more detail. First of all, one analyses the correct recovery of the ground state Wigner function~(\ref{wif0-gh}) from the obtained ground state expression (\ref{wigfg0-2}). One writes down the following expansions:
\begin{eqnarray}
 g_0 ^{2\lambda _0 ^2 a^2  + 1}  &\approx& e^{ - 2\lambda _0 ^2 x_0 \left( {x_0  - a} \right)} , \nonumber\\ 
 e^{ - 2g_0 \lambda _0 ^2 a\left( {x + a} \right)}  &\approx& e^{ - 2\lambda _0 ^2 a\left( {x + x_0  + a} \right) + \lambda _0 ^2 x_0 \left( {x_0  - 2x} \right)} , \nonumber\\ 
 \left( {m_0 \omega \frac{{x + a}}{p}} \right)^{\lambda _0 ^2 a^2  + \frac{1}{2}}  &\approx& \left( {\frac{{m_0 \omega a}}{p}} \right)^{\lambda _0 ^2 a^2  + \frac{1}{2}} e^{ - \frac{1}{2}\lambda _0 ^2 x\left( {x - 2a} \right)} , \nonumber 
\end{eqnarray}
as well as the following approximation and asymptotics:

\begin{eqnarray}
 \Gamma \left( {\lambda _0 ^2 a^2  + \frac{1}{2}} \right) &\cong& \sqrt {2\pi } \left( {\lambda _0 a} \right)^{2\lambda _0 ^2 a^2 } e^{ - \lambda _0 ^2 a^2 } , \nonumber\\ 
 J_{\lambda _0 ^2 a^2  + \frac{1}{2}} \left( {2\frac{p}{\hbar }\left( {x + a} \right)} \right) &\cong& \frac{1}{{\sqrt {2\pi } \lambda _0 a}}e^{ - \lambda _0 ^2 a^2  - \frac{{p^2 }}{{m_0 \omega \hbar }} - \frac{1}{2}\lambda _0 ^2 x\left( {x - 2a} \right) - \left( {\lambda _0 ^2 a^2  + \frac{1}{2}} \right)\ln \left( {2\lambda _0 ^2 a^2 } \right) + \left( {\lambda _0 ^2 a^2  + \frac{1}{2}} \right)\ln \left( {2\frac{a}{\hbar }p} \right)} . \nonumber 
\end{eqnarray}

Their substitution at (\ref{wigfg0-2}) with further substitution of the natural logarithm expansion under the case $a \to \infty$ yields:

\be
\label{wf0-lim}
\mathop {\lim }\limits_{a \to \infty } W_0^g \left( {p,x} \right) = W_{N0}^g \left( {p,x} \right).
\ee

We are going to prove the correct limit existence of the excited Wigner function through its finite-difference operator action version. There exists the following finite-difference operator action version of the Wigner function~(\ref{wif-gh}):

\be
\label{wfn-fd}
W_{Nn}^g \left( {p,x} \right) = \frac{1}{{2^n n!}}H_n \left( {\lambda _0 \left( {x + x_0  - \frac{{i\hbar }}{2}\frac{\partial }{{\partial p}}} \right)} \right)H_n \left( {\lambda _0 \left( {x + x_0  + \frac{{i\hbar }}{2}\frac{\partial }{{\partial p}}} \right)} \right)W_{N0}^g \left( {p,x} \right).
\ee

Then, one can rewrite (\ref{wigf-7}) also in its following finite-difference operator action version:

\be
\label{wfc-fd}
W_n^g \left( {p,x} \right) = \frac{{n!}}{{\left( {2\lambda _0 ^2 a^2  + 1} \right)_n }}L_n^{\left( {2\lambda _0 ^2 a^2 } \right)} \left( {2\lambda _0 ^2 g_0 a \left( {x + a - \frac{i\hbar}{2} \frac{\partial }{{\partial p}}} \right)} \right)L_n^{\left( {2\lambda _0 ^2 a^2 } \right)} \left( {2\lambda _0 ^2 g_0 a \left( {x + a+ \frac{i\hbar}{2} \frac{\partial }{{\partial p}}} \right)} \right)W_0^g \left( {p,x} \right).
\ee

Now, taking into account that there is the following known direct limit from the Laguerre to Hermite polynomials~\cite{koekoek2010}:

\[
\mathop {\lim }\limits_{\alpha  \to \infty } \left( {\frac{2}{\alpha }} \right)^{\frac{n}{2}} L_n^{\left( \alpha  \right)} \left( {\alpha  + \sqrt {2\alpha } z} \right) = \frac{{\left( { - 1} \right)^n }}{{n!}}H_n \left( z \right),
\]
one can easily show that the following correct limit from (\ref{wfc-fd}) to (\ref{wfn-fd}) also holds:

\be
\label{wfn-lim}
\mathop {\lim }\limits_{a \to \infty } W_n^g \left( {p,x} \right) = W_{Nn}^g \left( {p,x} \right).
\ee

In this paper, we have constructed the phase space representation for a semiconfined harmonic oscillator model with a position-dependent effective mass in terms of the Wigner quasiprobability function. We considered two cases with and without applied external homogeneous field and found that the Wigner distribution function for both cases is expressed through the finite sum of the product of the Bessel function of the first kind and the generalized Laguerre polynomials. The existence of the correct limits from these functions to their non-relativistic canonical analogues is proven, too. \cite{quesne2022} also generalizes a semiconfined harmonic oscillator model with a position-dependent effective mass introduced in~\cite{jafarov2021} and extends the model to the case of the so-called semiconfined shifted oscillator. Despite the extension, wavefunctions of the family of the semiconfined harmonic oscillator potentials preserve their general behavior in terms of the Laguerre polynomials. Therefore, the method developed here for the computation of the exact expression of the Wigner function can be further applied to semiconfined shifted oscillator potentials, too.

\section*{Acknowledgement}

This work was supported by the Azerbaijan Science Foundation –- Grant Nr \textbf{AEF-MCG-2022-1(42)-12/01/1-M-01}.

\section*{Data Availability Statement}

Data sharing is not applicable to this article as no new data were created or analyzed in this study.

\end{document}